\documentclass[preprint2]{aastex62}

\shorttitle{Explosion Mechanism of SNe Ia and Nucleosynthesis}
\shortauthors{Mori et al.}

\begin{document}

\title{Nucleosynthesis Constraints on the Explosion Mechanism for Type Ia Supernovae}

\author{Kanji Mori}
\affil{
     Graduate School of Science,
         The University of Tokyo, 7-3-1 Hongo, Bunkyo-ku, Tokyo, 
113-0033 Japan}
\affil{National Astronomical Observatory of Japan 2-21-1
             Osawa, Mitaka, Tokyo, 181-8588 Japan}
\email{kanji.mori@nao.ac.jp}
\author{Michael A. Famiano}
\affil{Department of Physics, Western Michigan University, Kalamazoo, 
Michigan 49008 USA}
\affil{National Astronomical Observatory of Japan 2-21-1
             Osawa, Mitaka, Tokyo, 181-8588 Japan}
\author{Toshitaka Kajino}
\affil{
     School of Physics and Nuclear Energy Engineering, and Internationsl 
Research Center for Big-Bang Cosmology and Element Genesis, Beihang 
University, Beijing 100083, P.R. China}
\affil{National Astronomical Observatory of Japan 2-21-1
             Osawa, Mitaka, Tokyo, 181-8588 Japan}
             \affil{
     Graduate School of Science,
         The University of Tokyo, 7-3-1 Hongo, Bunkyo-ku, Tokyo, 
113-0033 Japan}
\author{Toshio Suzuki}
\affil{
     Department of Physics, College of Humanities and Sciences,
     Nihon University 3-25-40 Sakurajosui, Setagaya-ku, Tokyo 156-8550 
Japan}
\affil{National Astronomical Observatory of Japan 2-21-1
             Osawa, Mitaka, Tokyo, 181-8588 Japan}
\author{Peter M. Garnavich}
\affil{
     Departmant of Physics, {Center for Astrophysics,} University of Notre Dame, Notre Dame, IN 
46556, USA}
\author{Grant J. Mathews}
\affil{
     Departmant of Physics, {Center for Astrophysics,} University of Notre Dame, Notre Dame, IN 
46556, USA}
\affil{National Astronomical Observatory of Japan 2-21-1
             Osawa, Mitaka, Tokyo, 181-8588 Japan}
\author{Roland Diehl}
\affil{
     Max Planck Institut f\"ur extraterrestrische Physik, D-85748 
Garching, Germany}
\affil{National Astronomical Observatory of Japan 2-21-1
             Osawa, Mitaka, Tokyo, 181-8588 Japan}
\author{Shing-Chi Leung}
\author{Ken'ichi Nomoto}
\affil{Kavli Institute for the Physics and Mathematics of the Universe (WPI), The University of Tokyo, Kashiwa, Chiba 277-8583, Japan}



\begin{abstract}
Observations of type Ia supernovae include information about 
the characteristic nucleosynthesis associated with these 
thermonuclear explosions. We consider observational constraints from 
iron-group elemental and isotopic ratios,
to compare with various  models  obtained with the most-realistic 
recent treatment of electron captures.
The nucleosynthesis is sensitive to  the  highest white-dwarf central 
densities.  Hence, nucleosynthesis yields can distinguish 
high-density Chandrasekhar-mass models from lower-density burning models such as white-dwarf  mergers.
We discuss {new} results of 
post-processing nucleosynthesis {for two spherical models  (deflagration and/or  delayed detonation models) based upon new electron capture rates}.
{We also consider cylindrical} and  3D explosion models (including  deflagration, delayed-detonation, or a violent merger {model}).  
{Although there are} uncertainties in the observational constraints, we identify some trends in observations and the models. We {make a new comparison of} the models with elemental and isotopic {ratios} from five observed supernovae and three supernova remnants.  We find that the models and data tend to fall {into} two groups.  In one group low-density cores such as in a 3D merger model are {slightly more} consistent with the nucleosynthesis data, while the other group is {slightly better} identified with higher-density cores such as in single-degenerate 1D-3D deflagration models. Hence, we postulate that both types of environments {appear to} contribute nearly equally to observed SNIa.
We also note that observational constraints on the yields of $^{54}$Cr and $^{54}$Fe, if available, might be 
used as a means to clarify the {degree of geometrical symmetry} of  SNIa explosions.
\end{abstract}
\section{Introduction}
\label{intro}
Type Ia supernovae (SNIa) are thought to result from accreting CO or white dwarfs (WDs) in close binaries
 \citep[e.g.,][]{hoyle60,arnett96,hillebrandt00,boyd08,illiadis12}.  For sufficiently high  central densities, the { initiation of} carbon, oxygen, and/or neon thermonuclear burning in the core
can result in a violent explosion {that disrupts} the entire star. The subsequent nucleosynthesis can result in an abundance of
Fe-peak elements. The ejection of these elements into the interstellar medium (ISM) contributes to the galactic chemical enrichment.
Moreover, a carefully-selected subset of SNIa are currently employed as standard candles in cosmology to measure the {acceleration} of the Universe \citep{riess98,perlmutter99,schmidt08}.

Despite their importance in Fe-peak enrichment in the ISM, the origin of SNe Ia, including the progenitors and the actual explosion mechanism is still a subject of 
debate
\citep[e.g.,][]{maoz14,hillebrandt00,nomoto95}.
Two major progenitor models have been hypothesized.  One case involves
accretion from a non-degenerate companion. The WD mass
then approaches the Chandrasekhar mass inducing a SN Ia.  This is known as the single-degenerate progenitor model and the 
accreting WD is dubbed a ``Chandra model'' star because
its mass approaches the Chandrasekhar mass. The other case is the double-degenerate model.  This case involves two 
sub-Chandrasekhar-mass WDs (``sub-Chandra'') that merge to
form a SN Ia
\citep{iben84,webbink84}. {Several} violent merger (VM) models \citep[e.g.,][]{pakmor,sato16} have 
 {attempted to simulate} these explosive events.

The Chandra and sub-Chandra models involve different  central
densities $\rho_c$ of the white dwarf at the time of central ignition.  In the Chandra model,
$\rho_c > 10^9$ g cm$^{-3}$, while $\rho_c \lesssim 10^8$ g cm$^{-3}$
in the sub-Chandra models \citep{wang12,nomoto11}.  This 
changes the explosion dynamics of the subsequent
supernova.  
For Chandra models, the thermonuclear burning propagates
outward as a subsonic flame front known as a deflagration wave
\citep{nomoto76,nomoto84}.  Burning is expected to be enhanced by an
increase in the surface area from Rayleigh-Taylor instabilities
at the front 
\citep{muller82,arnett94,khokhlov95}.  This front may undergo
a deflagration-to-detonation transition
 \citep[DDT:][]{blinnikov86,khokhlov91,nomoto} for a strong
 enough deflagration.
This type of explosion dynamics has been modeled in 3D simulations
\citep[e.g.,][]{gamezo05,roepke06,seitenzahl13}.

The explosion mechanism of SNe Ia is also related to its 
nucleosynthesis. The results of this nucleosynthesis may be
inferred from  direct light-curve observations and/or spectral observations
of the remnants. {Indeed}, the decays of \(^{56}\)Ni and its daughter \(^{56}\)Co are the primary power source of 
the light-curves \citep{arnett79}.  

For a sufficiently high central density the Fermi energy
of the electrons can exceed the energy threshold for electron capture (EC) reactions on ambient nuclei.
An increase in the EC rates in the subsequent nucleosynthesis, can reduce the overall
electron fraction $Y_e$, defined as the sum over all nuclear
species:
\begin{equation}
Y_e \equiv \sum_i Z_i Y_i,
\end{equation}
where $Y_i$ is the abundance of a given species with proton
number $Z_i$.
Thus, the larger central densities associated with the Chandra
models are expected to  result in a shift to lower electron fraction and
a shift of the nucleosynthesis toward more neutron-rich nuclei as
compared to the sub-Chandra models. 
However, this may be influenced by a variety of features in the models such as the flame speed, convection, and
rotation of the system \citep{benvenuto15}. 

{The new aspects of the present work are to make revised nucleosynthesis calculations based upon new EC rates. We also make a new summary of the nucleosynthesis products in several recently observed  light curves.} 

In this paper we have selected observational constraints  for the purpose of 
comparing specific nucleosynthesis {predictions} of SNIa models. Toward that aim we
make use of our recently updated nucleosynthesis treatment  
concerning the weak reaction rates involved in electron captures and the 
resulting change in $Y_e$. {The important roles of  nuclear e-capture rates on $pf$-shell nuclei for the synthesis of iron-group elements in SN Ia explosions have been discussed in \citep{langanke03}, where  the KBF and KB3G shell-model Hamiltonians were used. However, new e-capture rates have been obtained by \citep{honma17} with a new $pf$-shell Hamiltonian, GXPF1J, which can better describe the Gamow-Teller (GT) strengths in Ni and Fe isotopes. Especially in $^{56}$Ni, the GXPF1J can reproduce the experimental GT data very well in contrast to  the KB3G and KBF.  A comparison of the GT strengths and e-capture rates in $^{56}$Ni and $^{54}$Fe as well as the impact of the new rates on elemental abundances in SN Ia explosions are discussed  in detail in \citep{mori16}.} 

 Therefore, in this paper we concentrate on iron group elemental and 
isotopic {ratios} as
observational constraints. We then discuss a  number of selected models intended to 
address the  range of nucleosynthesis conditions relevant to those iron group elements and isotopes. We include 
both Chandra- and sub-Chandra-models here, and also consider  1D, 2D,  and 3D 
simulations.  In this way  we can test whether the {assumed symmetry} of the models or the 
type of explosion {mechanism} are the main features characterizing the resultant nucleosynthesis. 

We find that the models and observed nucleosynthesis tend divide  into two nearly equally favored groups roughly characterized by the central densities during thermonuclear burning.  One set of data is consistent with the low central densities of sub-Chandra merger models, while the other favors the higher central densities of Chandra models.  We also speculate on which  future 
observables might help to distinguish {the degree of symmetry in these}  explosions.

Section \ref{observations} presents the observational constraints that we {have} selected for
our comparisons among models. Section \ref{exp_model} presents the {available models 
covering} the space of hypothetical nucleosynthesis conditions. 
In Section \ref{results} we compare various models both {with} each other and with the observational 
constraints.   Section \ref{discussion} presents the conclusions  and provides suggestions for future 
work.

\section{Observational Constraints}
\label{observations}
Supernova light curves and their spectral evolution have been observed for many events and {have been} frequently discussed in the context of  theoretical models \citep[see, e.g.,][]{hoflich17}.
Here, we select observations of recent years, from which key nucleosynthesis results are extracted to compare to  nucleosynthesis results in the models. Observations mainly focus on 
\begin{itemize}
\item[(1)] The late-time light curve evolution and its inferred energy source.
\item[(2)] Supernova-remnant abundances for iron group elements including Mn.
\item[(3)] Direct {$\gamma$-ray} measurements of {the} ejected mass of $^{56}$Ni.
\item[(4)] A comparison to solar abundances, specifically for Mn.
\end{itemize}
\par\noindent
A summary of the observational data employed in the present study is summarized  in 
Table \ref{wd_obs}.

	\begin{deluxetable*}{c|ccccccccc}
\tabletypesize{\scriptsize}
		\tablecaption{\label{wd_obs}Observations}
		\tablehead{  & \colhead{SN 2011fe [1]} & \colhead{SN 2012cg [2]} & \colhead{SN 2014J [3,4]} &\colhead{SN 2015F [5]} &\colhead{SN 2013aa [6]}&\colhead{SNR 3C 397 [7]}& \colhead{Kepler [7]} & \colhead{Tycho [7]} & \colhead{Solar [8]} }
		\startdata
		$\log_{10}(^{57}\mathrm{Co}/^{56}\mathrm{Ni})$ & $-1.59^{+0.06}_{-0.07}$ &$-1.36^{+0.11}_{-0.13}$ & $-1.18^{+0.06}_{-0.05}$&$-2.40^{+0.25}_{-0.30}$&$-1.70^{+0.18}_{-\infty}$ \\
		$\log_{10}(^{55}\mathrm{Fe}/^{57}\mathrm{Co})$& $-1.1^{+0.2}_{-0.4}$ & ~ &  \\
		$M_{^{56}\mathrm{Ni}}$ [$M_\odot$] & $0.50(2)$&  $\approx0.7$&$0.49(9)$\\
		$M_{^{57}\mathrm{Co}}$ [$M_\odot$] & 0.012(2)&&\\
		Mn/Fe&&&&&&$0.025^{+0.008}_{-0.007}$ & $0.010^{+0.007}_{-0.0035}$ & $0.013^{+0.007}_{-0.005}
$ & $0.0084(10)$\\
		Ni/Fe&&&&&&$0.17^{+0.07}_{-0.05}$ &$ 0.045^{+0.03}_{-0.015}$&$0.025^{+0.015}_{-0.01}$&0.054(7) 
		\enddata
		\tablecomments{[1] \cite{shappee17}; [2] \cite{graur16}; [3] \cite{diehl15}; [4] \cite{2014j}; [5] \cite{2015f}; [6] \cite{2013aa}; [7] \cite{yamaguchi15}; [8] \cite{asplund09}}
	\end{deluxetable*}

\subsection{Light-curve analysis}
The usefulness of the late-time light-curve analysis with respect to the presence or absence of longer-lived radioactive energy sources has been demonstrated for SN1987A \citep{fransson02}, and was discussed  by \citet{,seitenzahl09} in the context of  various isotopes such as $^{55}$Fe. {It has also been pointed out that charactaristic X-rays from long-lived nuclei can be used to study supernova nucleosynthesis \citep{leising01}.}

Recently, key results have been added for SN 2011fe {\citep{shappee17},  SN 2012cg \citep{graur16}}, {SN 2014J  \citep{2014j}, SN 2015F \cite{2015f}, and SN 2013aa \citep{2013aa}}. 
\citet{shappee17} used HST and the Large Binocular Telescope to follow the light curve of nearby SN2011fe for an unprecedented time of 1840 days. In particular, they tested  for the decays of $^{57}$Co$\rightarrow$$^{57}$Fe ($t_{1/2}=271.79$ d) and $^{55}$Fe$\rightarrow$$^{55}$Mn ($t_{1/2}=999.67$ d). 
The late-time light curve of SN 2011fe is fit significantly better if the radioactive-energy input from $^{57}$Co and $^{55}$Fe  are included. 
Their best fit was for abundance ratios of $^{57}$Co/$^{56}$Ni = 0.03,  and $^{55}$Fe/$^{57}$Co = 0.07. 
The fit is also acceptable without the contribution from $^{55}$Fe. {Hence, they only} provide a  2$\sigma$ upper limit  of $^{55}$Fe/$^{57}$Co $\le 0.22$.  
{Nevertheless, this inferred abundnace of $^{57}$Co is consistent with the direct detection of $\gamma$-rays from $^{57}$Co in SN 1987A \citep{kurfess92}.} We use their estimate for SN2011fe of $\log(^{57}\mathrm{Co}/^{56}\mathrm{Ni})=-1.59^{+0.06}_{-0.07}$.

A recent late-time (day 1034) spectrum of SN2011fe \citep{taubenberger} supports this indirectly: \cite{fransson15} find the need for energy injection by
$^{57}$Co. {Also}, from the observed flux level in their spectrum they require a production ratio of $^{57}$Ni/$^{56}$Ni of 2.8 times the solar ratio.
\citet{shappee17}  concluded that a violent merger model is favored for SN2011fe because it produces a $^{55}\mathrm{Fe}/^{57}\mathrm{Co}$ ratio of 0.27, which is in better agreement with the observational limit. {However, it should be noted that the estimated isotopic ratio is subject to systematic uncertainties due to a lack of detailed knowledge on the physical processes including electron/positron escape and light echoes \citep{kerzendorf,dimitriadis}.}

\citet{graur16} also used HST to track the light curve of SN 2012cg out to 1055 days. They determined the slope of the light curve decay at early times when $^{56}$Co decay powers the light curve and the supernova should already be transparent to gamma rays.  They then extrapolated to times beyond day 500 to find that they needed another source of energy. 
A blue excess that had been reported in the early light curve of SN 2012cg could be interpreted as evidence for an interaction between the supernova ejecta and a non-degenerate companion \citep{marion16}. This interpretation is debated by \citet{shappee16} from a re-analysis of these data. Contamination by the light echo for this supernova is excluded by \citet{2015f}.
From these considerations, \citet{graur16} estimated a mass ratio $^{57}$Co/$^{56}$Co of 0.043$^{+0.012}_{-0.011}$ or
 $\log(^{57}\mathrm{Co}/^{56}\mathrm{Ni})=-1.36^{+0.11}_{-0.13}$. 
This is somewhat higher than the ratio determined for SN 2011fe. But we note that such constraints are indirect, and depend upon assumptions about how the light-curve evolution is driven by those radioactive species.

{In similar ways, the $^{57}$Co/$^{56}$Ni ratio has been estimated for SN 2014J, 2015F, and 2013aa using HST. \cite{2014j} observed late light curves of SN 2014J to 1181 days and estimated $^{57}\mathrm{Co}/^{56}\mathrm{Ni}=0.066^{+0.009}_{-0.008}$, which is higher than the other objects. On the other hand, \citep{2015f} observed light curves of SN 2015F to 1040 days and estimated $^{57}\mathrm{Co}/^{56}\mathrm{Ni}=0.004^{+0.003}_{-0.002}$, which is significantly lower than the others. \cite{2013aa} observed light curves of SN 2013aa to 1500 days and reported $^{57}\mathrm{Co}/^{56}\mathrm{Ni}=0.02^{+0.01}_{-0.02}$, although this result is consistent with zero.}

\subsection{SNIa  Remnants}
Observations of characteristic X-ray recombination lines in supernova remnants allow constraints on the production of new nuclei in supernovae,  even though {one must correct for the fact that} some of the radiating material will be swept up from the surrounding gas. \citet{yamaguchi15} { analysed the Type Ia supernova remnant (SNR) 3C 397  with archival data of the Suzaku X-ray mission.} From the observed K-shell emission lines they inferred an excess in Ni and Mn production with respect to Fe, with $\mathrm{Ni}/\mathrm{Fe}=0.17^{+0.07}_{-0.05}$ and $\mathrm{Mn}/\mathrm{Fe}=0.025^{+0.008}_{-0.0077}$. This is indicative that burning at higher density may have occurred in this explosion, suggesting a Chandrasekhar-mass explosion.

\subsection{Mass of $^{56}$Ni}
It is common practice to estimate the $^{56}$Ni mass produced in {SNIa explosions}  through Arnett's rule, i.e. one interprets the brightness at maximum light in terms of the radioactive input from $^{56}$Ni decay.
For the first time, however, 
\citet{diehl15} directly detected and traced the brightness of the characteristic  $\gamma$-ray lines at 847 and 1238 keV from the $\beta$-decay chain of $^{56}$Ni in SN 2014J. Comparing light curves for a variety of 1D models, they derived a $^{56}$Ni mass of $(0.49\pm0.09)M_\odot$. The quoted uncertainty includes the range of model dependence, as models must be used to {interpret} the characteristic brightness evolution of the gamma-ray emission at the time the supernova gradually leaks out these gamma rays. 

From a similar analysis of the same INTEGRAL data, \citet{churazov15}  inferred a Ni mass constraint of 0.54-.67 $M_\odot$, using a different set of models for the gamma-ray light curve, and fixing the poorly constrained ejecta mass to 1.4 $M_\odot$. Obviously, even for such rather direct measurements, model dependencies still contribute to the uncertainties.
Nevertheless, the $^{56}$Ni mass estimate is consistent with those {deduced from} Arnett's rule and optical data, and also the $^{56}$Ni production in SN 2011fe estimated by \cite{shappee17}. We choose the value and uncertainties of $(0.49\pm0.09)M_\odot$ for our analysis here.

\subsection{Solar abundances}
The abundances of elements in solar material also provide an observational constraint, as nucleosynthesis in AGB and massive stars and their core collapse supernovae have been shown not to produce Mn in excess of the solar ratios \citep{nomoto15}. 
{\cite{seitenzahl13b} and \cite{kobayashi15} compared the Mn production of various SNIa models with the solar Mn/Fe ratio. The Mn production is sensitive
to the progenitor mass (near-Chandra or sub-Chandra) and only near-Chandra models predict $[\mathrm{Mn}/\mathrm{Fe}]>0$. They concluded that [Mn/Fe] in the solar system cannot be reproduced without near-Chandra models, while the best fit is for equal portions of near-Chandra and sub-Chandra models. } We independently deduce a similar conclusion here based upon  the observed nucleosynthesis.

\section{Explosion Models}
\label{exp_model}
The various explosion models considered  in the present work are summarized in Table \ref{wd_calculations}.  These span the gamut of {symmetry from spherical to 3D and from} Chandra to sub-Chandra environments.  Among these models, in the present work we have 
explicitly re-evaluated the nucleosynthesis of two explosion models in a post-processing network.
Trajectories from the W7 
deflagration \citep{nomoto84,thielemann86}
and the WDD2 delayed detonation models  \citep{nomoto} were re-evaluated in an updated  nucleosynthesis network.
Nucleosynthesis results were then compared to those from other explosion models and to observations of SNe Ia light-curve data and remnants.

The W7 deflagration model is a 1D 
flame propagation model.  The explosion is postulated to result in an increase of nuclear burning 
due to an increase in the surface area of the flame front as
it propagates outward toward the surface of the white dwarf.  In this 
explosion mechanism, the speed of the flame is 
derived from mixing-length theory.  {In the W7 model \citep{nomoto84},
the flame accelerates to 0.08$c_s$   after 0.6 s and to 0.3$c_s$ after 1.18 s (where $c_s$ is the local sound speed).}

In the case of the WDD2 delayed detonation model \citep{nomoto},
the explosion begins as a {spherical} deflagration and transitions to
a detonation at a low density \citep{khokhlov91} with the transition 
density given as a parameter of the model.  
In the model studied here, the deflagration phase 
was taken from the description and parametrisation of \cite{nomoto84} and was
followed by the detonation phase taken from the numerical 
results given by \cite{nomoto} with a transition 
density at 2.2$\times10^7$ gcm$^{-2}$.  In the present work, the nucleosynthesis was {re-calculated} for the spherically symmetric trajectories
of the WDD2 model.  

To evaluate the differences in
nucleosynthesis in each explosion, we ran a nuclear reaction network \citep{libnucnet} decoupled from the explosion model
using individual trajectories from each explosion as input to the network. Each
trajectory was obtained as a  sequence of burning at time, $t$, in a particular mass range.  At each time step, the electron chemical potential and electron fraction
$Y_e$ were computed implicitly for the {determination} of the electron capture (EC) rates. The reaction network used the EC rates computed from the GXP shell model 
(\cite{honma04}, \cite{honma05}, \cite{suzuki11}) for the $pf$-shell nuclei 
with 21$\leq Z\leq$ 32 and mass $A$, $42\leq A\leq 71$. These rates have 
been updated from those used in \cite{mori16} by extending the region of 
$pf$-shell nuclei from $45\leq A\leq65$ as well as including the 
back-resonance contributions \citep{langanke01}. 
They are given for 
densities $\rho Y_e= 10^{5}\sim10^{11}$ g/cm$^{3}$ and temperatures 
$T_9 = 10^{-2}\sim10^{2}$ with $T = T_9\times 10^{9}$ K \citep{honma17}. 
The rates of \cite{oda} were used for
$sd$-shell nuclei and the rates of \cite{ffn1,ffn2} were used for the remaining rates.  
Rates for all other reactions were taken from the JINA REACLIB {V2.2} database \citep{reaclib}.

Some differences in the nuclear yields
are obtained for $^{65}$Cu in the W7 model. 
Apparent changes in the mass 
fractions (Table 3) of the neutron-rich Ca and Zn 
isotopes are minor and insignificant. 
The abundance of $^{41}$K is enhanced by 2-3 orders of magnitude compared to \cite{mori16} in both the W7 and the WDD2 model. This reflects the proper inclusion of the $\beta$ decay of $^{41}$Ca to $^{41}$K in this paper.
 Therefore,  the values reported here supersede those previously reported in \citep{mori16}.

We point out that our approach of post-processing does not account for the {second-order effect of the} different amount of heating from differences in the {nuclear} rates. This limitation of decoupling the
reaction network from the explosion trajectories, however, allows for a rapid evaluation of the nucleosynthesis. These approximations and their impact have been discussed in prior work \citep{mori16}.

{A comparison of the observed nucleosynthesis (cf. Table 1) to the calculated yields
among different models (cf. Table 2) 
 can provide insight into the explosion mechanism of SNIa.}
 The results of our {new spherical} network calculations are compared with the results of the {cylindrically symmetric} delayed detonation and deflagration models of \cite{leung18} as well as the 3D {(N100)} delayed detonation model  of \cite{seitenzahl13}, the 3D  {(N150def)} deflagration model of \cite{fink14}, and the { violent merger model of \cite{pakmor12}.} 
 Among  these, the N100 model is a three dimensional non-rotating, delayed detonation model with an initial central 
 density of 2.9$\times$10$^9$ g/cm$^3$, a mass of 1.4M$_\odot$, and an initial radius of 1.96$\times$10$^8$ cm. The N150def model is a 3D model for a deflagration
 scenario.  
 
 {Both of the 3D models were run with the same hydrodynamics codes. 
 The details of the computational methods are described in their respective references.  
 The violent merger model is a 3D simulation of two sub-Chandrasekhar mass WDs.  
 Our comparison among the W7, WDD2, LN18(def.), LN18(del.~det.), N100, N150def, and violent merger model 
 can shed light on the viability of each explosion model and effects of modeling in spherical, cylindrical, or unconstrained symmetry.}
 
 \section{Results discussion}
 \label{results}
 \subsection{Comparisons among models}
\begin{figure*}[htbp]
\plottwo{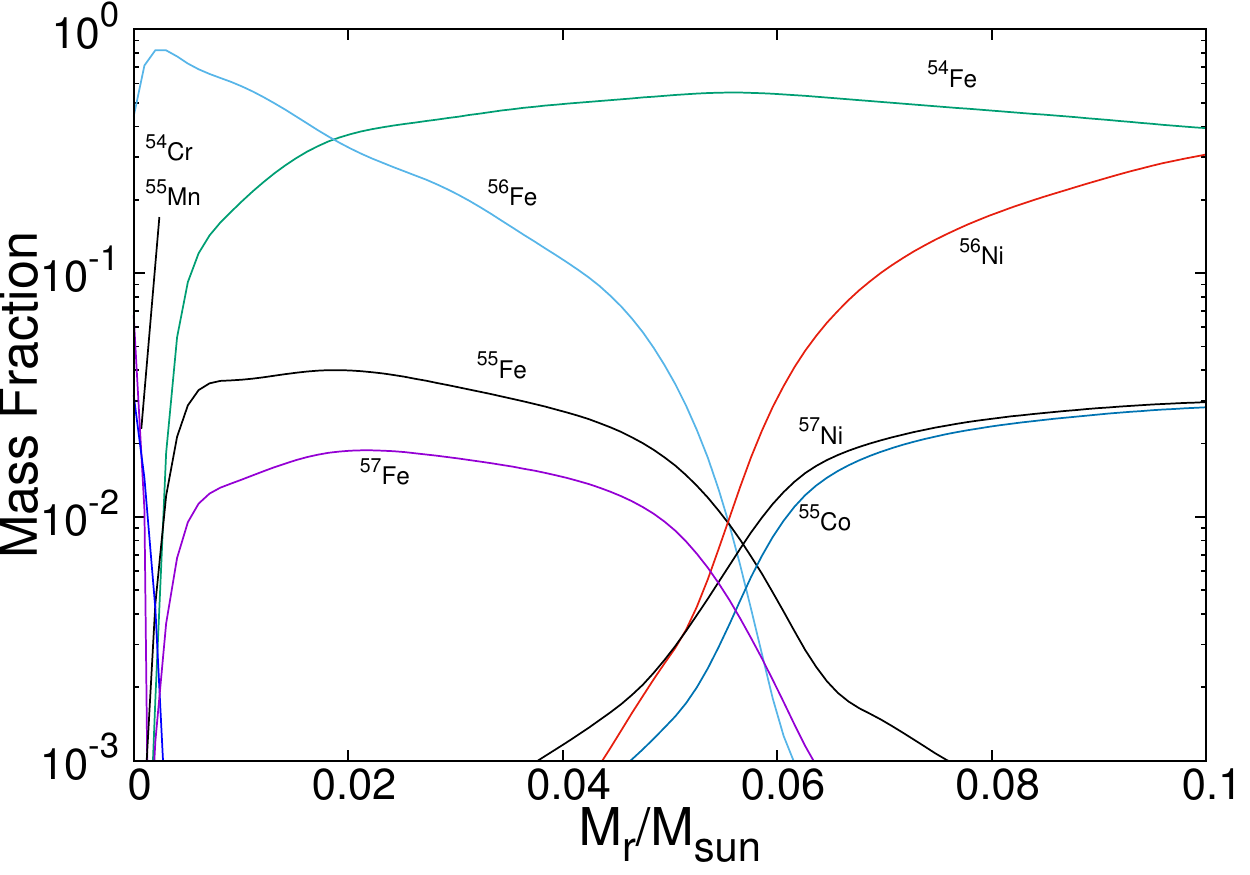}{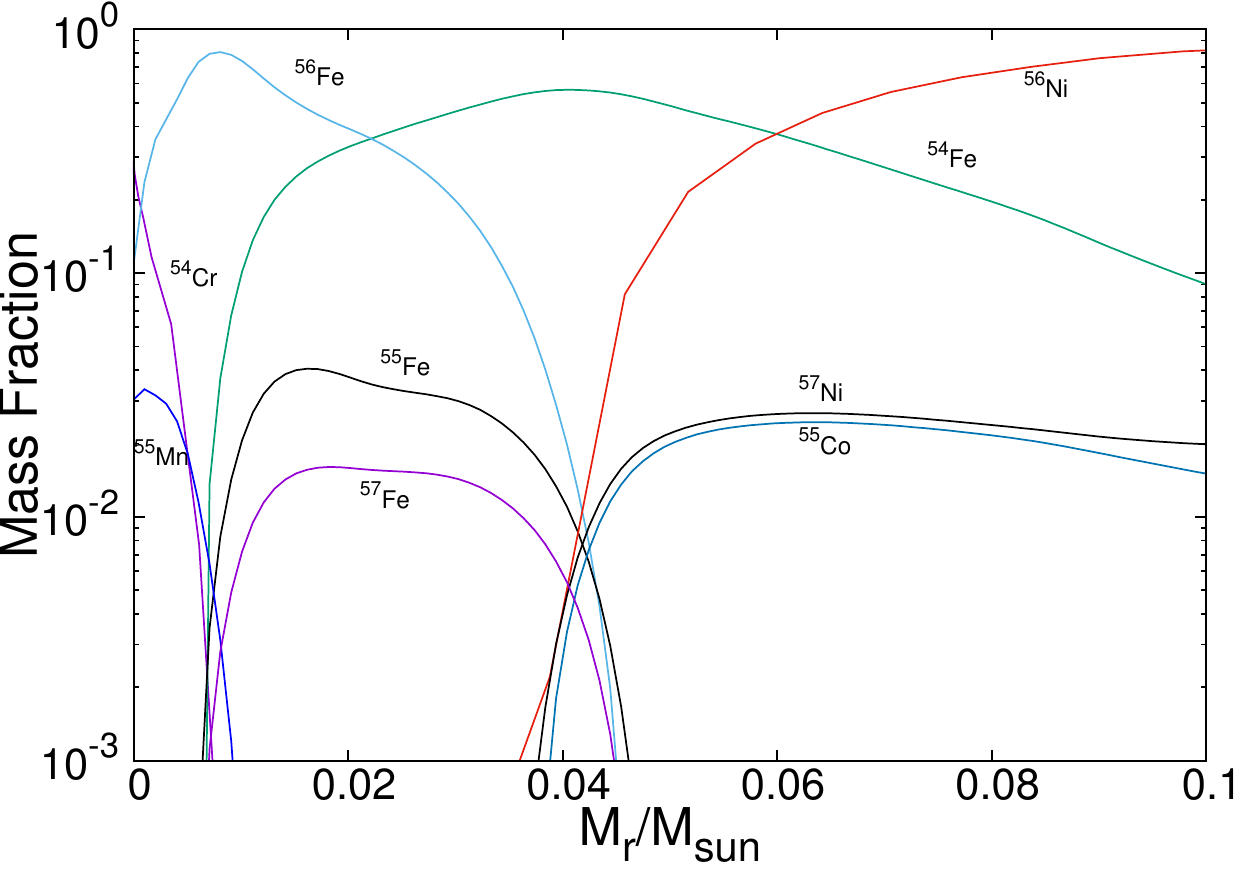}
\caption{\label{profiles}Abundance plots for the inner part of  the W7 model (left panel) and the WDD2 model (right panel). Only nuclei which are mentioned in the text are shown.}
\end{figure*}

 The final isotopic 
 abundance profiles for the updated W7 and WDD2 models computed in this work are
  shown in Figure \ref{profiles} as a function of ejected mass 
 for the inner 0.1 M$_\odot$ of the ejecta.
 The resulting mass fractions closely resemble those of prior work \citep{brachwitz00}. 

\begin{figure*}[htbp]
\plottwo{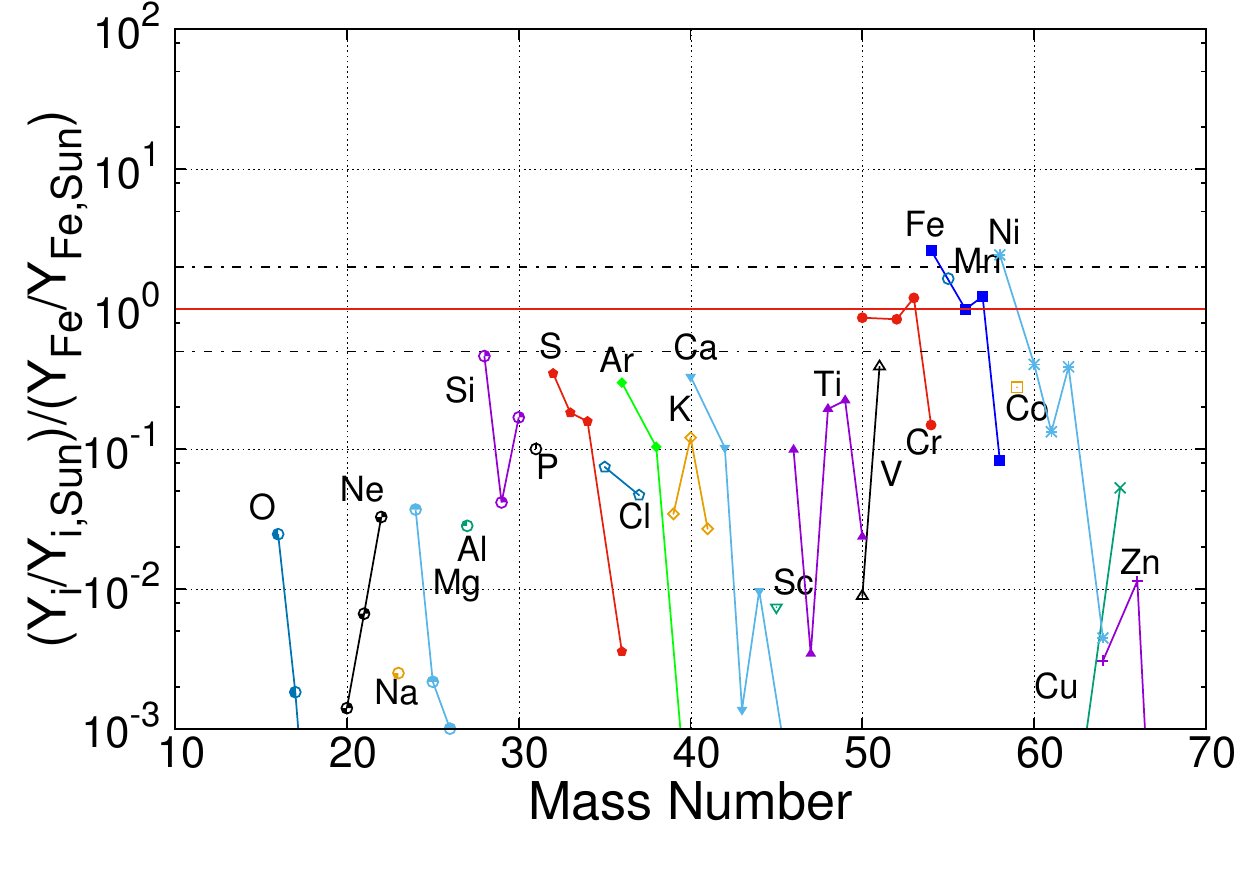}{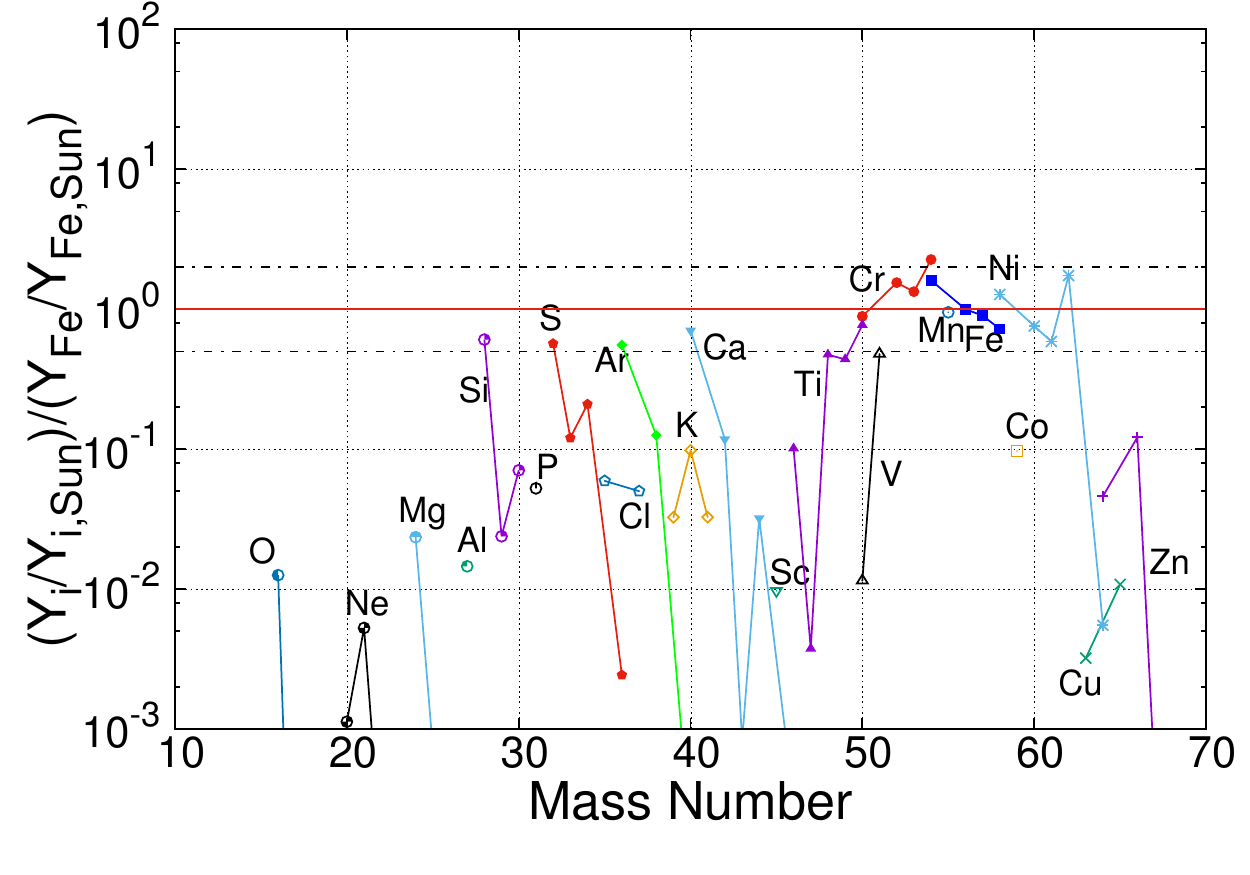}
\caption{\label{overproduction}Abundances of nuclei produced in the W7 model  (left) and the WDD2 model (right), 
normalized by the solar abundance and the $^{56}$Fe abundance \citep{mori16}.}
\end{figure*}

In addition to the isotopic abundance profiles, the total overproduction factors for the W7 and WDD2 explosion models 
are shown in Figure \ref{overproduction}.  This is defined by the ratio of
an isotopic species $i$ relative to iron, normalized to the same ratio in the solar abundance standard:

\begin{equation}
\frac{Y_i/Y_{Fe}}{Y_{i,\odot}/Y_{Fe,\odot}} 
	= \frac{Y_{i}/Y_{i,\odot}}{Y_{Fe}/Y_{Fe,\odot}}
\end{equation}
These abundances are the total ejected abundance in an event, calculated by weighting the yield in each shell by the mass of that shell.  

Both these models and their yields (evaluated with a different nuclear network) have been discussed previously 
\citep{brachwitz00}.  As we discussed in a previous paper \citep{mori16}, the WDD2 model appears to reproduce solar
distributions more closely for Cr, Mn, Fe, Ni, Cu, and Zn than the W7 model.  
Both of these models underproduce Cu and Zn, and show an overall trend of
enhanced production of lighter nuclei within an isotopic chain, with some exceptions.

The most notable of these exceptions is Cr: 
In the WDD2 model, neutron-rich Cr is enhanced relative to 
the proton-rich species in the chain, as compared to Cr production in the W7 model. 
Similarly, Cr is more
strongly produced in the central trajectories of the WDD2
explosion model, as seen in Figure \ref{profiles}.  In the
central 0.01 M$_\odot$ region of the star, the WDD2 model obtains a significant
mass fraction of Cr, while the W7 model predicts little Cr production.  
Hence, observational constraints on Cr could be a good indicator of the
explosion mechanism for a SN Ia event.

\begin{figure}[htbp]
\plotone{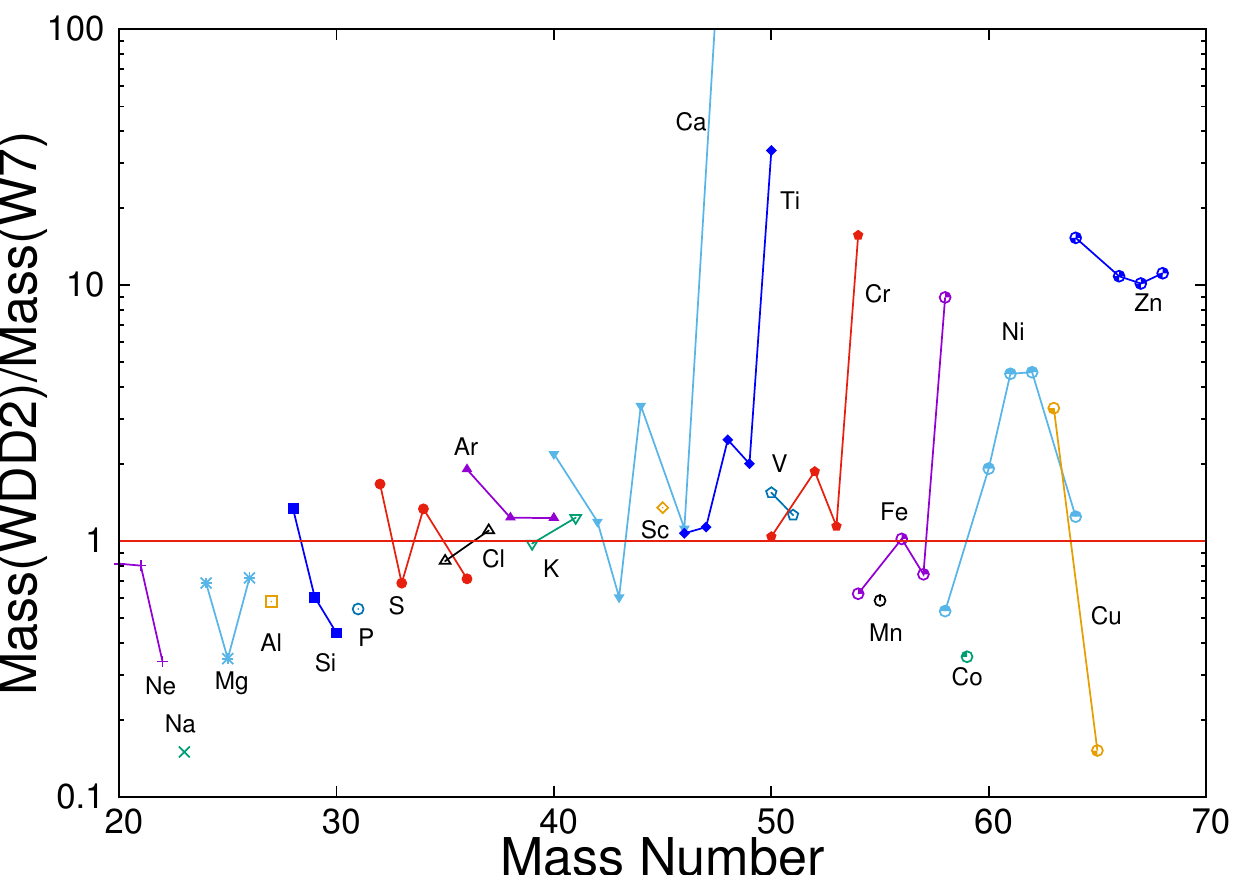}
\caption{\label{prod_ratio}Ratio of the ejected mass of nuclei produced in the WDD2 model to the mass of those produced in the W7 model \citep{mori16}.}
\end{figure}

A comparison of the W7 and WDD2 two models calculated in this work is summarized in Figure 
\ref{prod_ratio}, and the yields of several isotopes for these two models are shown in Figure \ref{all_yields}. 
The ratio of total mass yields for the WDD2 model to that of the W7 model is shown for various isotopic chains.  
For elements with $20\le Z\le 28$, there
is a general trend towards an increased production of neutron-rich nuclei (higher $Y_e$) in the WDD2 model.  The 
ratio of $^{54}$Cr is also noted as this is produced in reasonable abundance in the WDD2 model, but significantly
less in the W7 model.

Model results for various isotopic and elemental ratios are summarized in Table \ref{wd_calculations}.
These include the total mass fraction ratios $^{57}$Co/$^{56}$Ni and $^{55}$Fe/$^{57}$Co as well as the
total computed ejected mass of $^{56}$Ni, $^{57}$Co, $^{54}$Cr,
and $^{54}$Fe. In addition, the ratios of the total elemental mass of Mn/Fe and Ni/Fe are given.

The total mass of $^{56}$Ni ejected agrees within about 10\% for all models, except for the deflagration models LN18(def.) and N150def models, which predict a much lower yield.  
For $^{57}$Co, the W7, the LN18(del.det.)  and the N100 models appear to predict similar values, while the other models predict about 30\% less. 
The  $^{54}$Cr  yields, however,  differ dramatically among the models; this is a further indication that   this isotope as a good diagnostic 
for the explosion mechanism.
To a lesser extent, the production of $^{54}$Fe may constrain the explosion mechanism, as the values for similar models can vary by 
as much as 25\% from the average value.
The elemental ratios of Mn/Fe and Ni/Fe  may also be
used as indicators. The W7, LN18,  and N100 models predict similar ratios, while the WDD2 model predicts 
significantly less Mn/Fe and Ni/Fe, and the N150def model predicts somewhat larger ratios for  Mn/Fe and Ni/Fe.

{As an estimate of the} model uncertainties, we have made use of a recent broad parameter study for the LN18 SNIa models of \cite{leung18}.
In that study {cylindrically symmetric 2D} models were run {for} a wide range of central masses and composition, {along with different} flame ignition and propagation treatments. 
We use this broad study as a reference  {with fixed solar metallicity}, and assume that the other models have the same order of uncertainty of their yields.
That is, for each of the diagnostic ratios of interest here we take the extreme values from the range of models considered in \cite{leung18} as {an estimate of} the range of uncertainty in the computed results. We note however that {the \citet{leung18} models are Chandra deflagration and delayed-detonation models. As such they have discussed uncertainties related to for example the flame propagation, DDT, or ignition setup and density. However, the violent merger model does not share any of these uncertainties, as it consists chiefly of a detonation, with composition and mass of the primary as the most important parameters. Hence, we do not draw error bars for the violent merger model.}

\begin{figure*}[htbp]
\plottwo{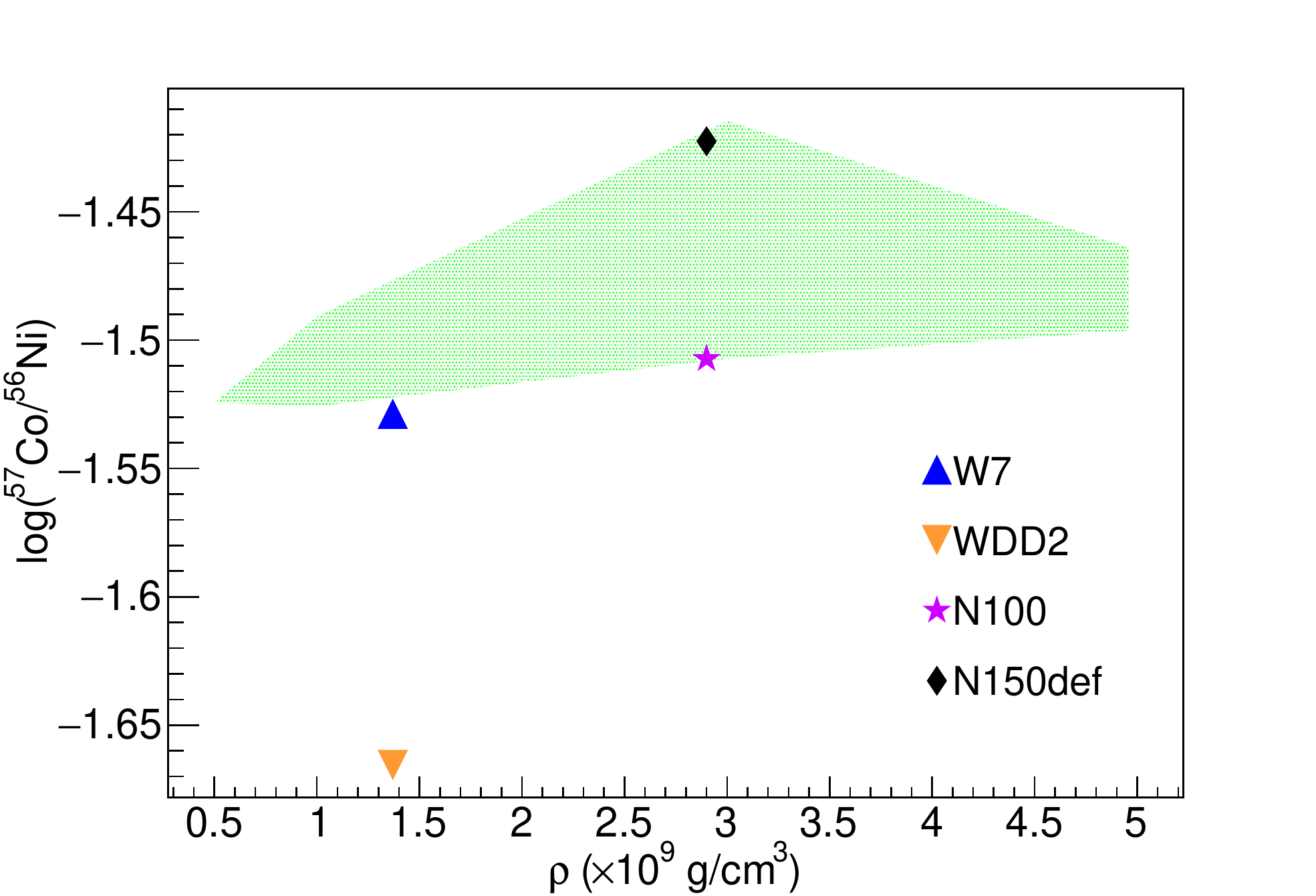}{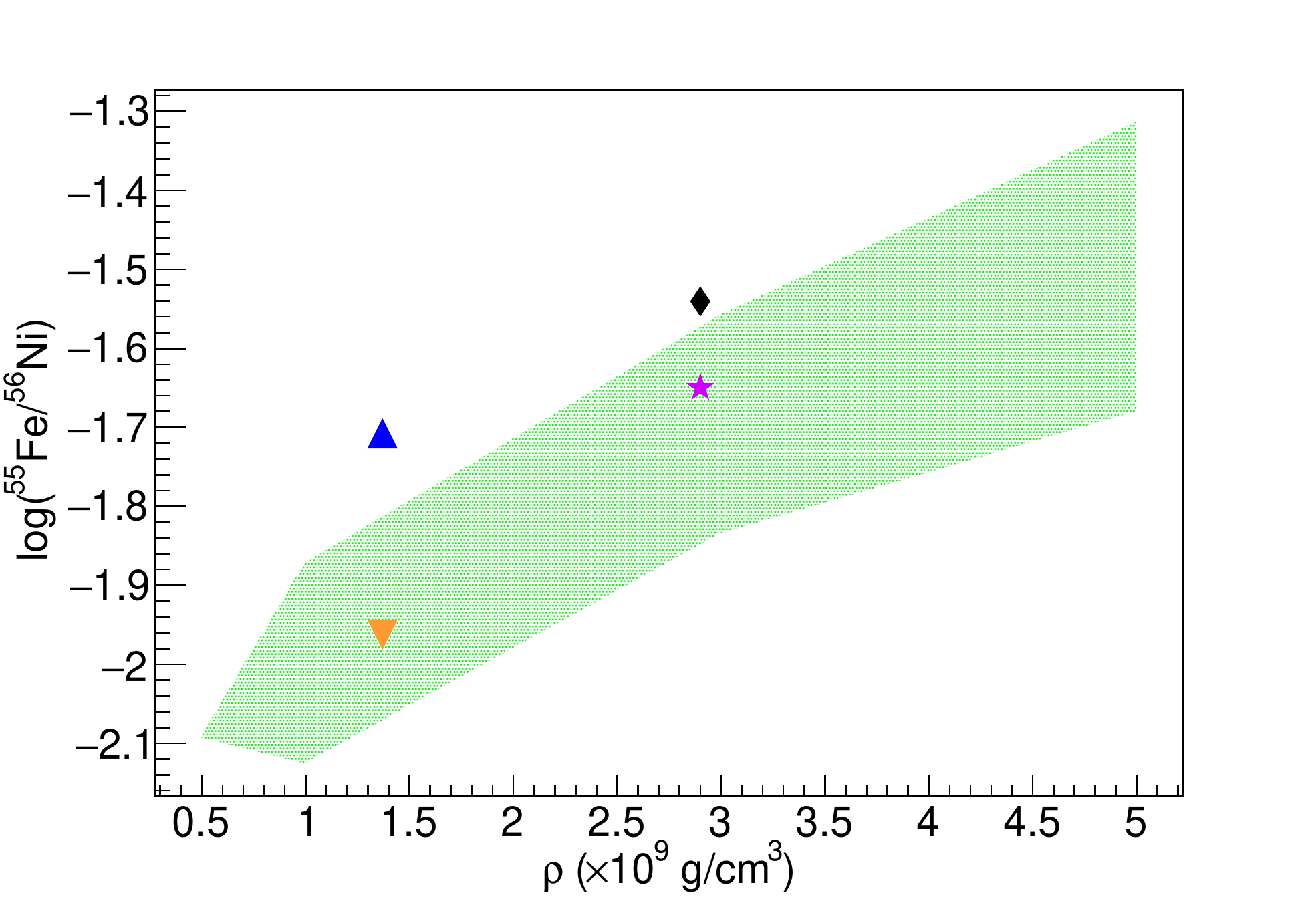}
\plottwo{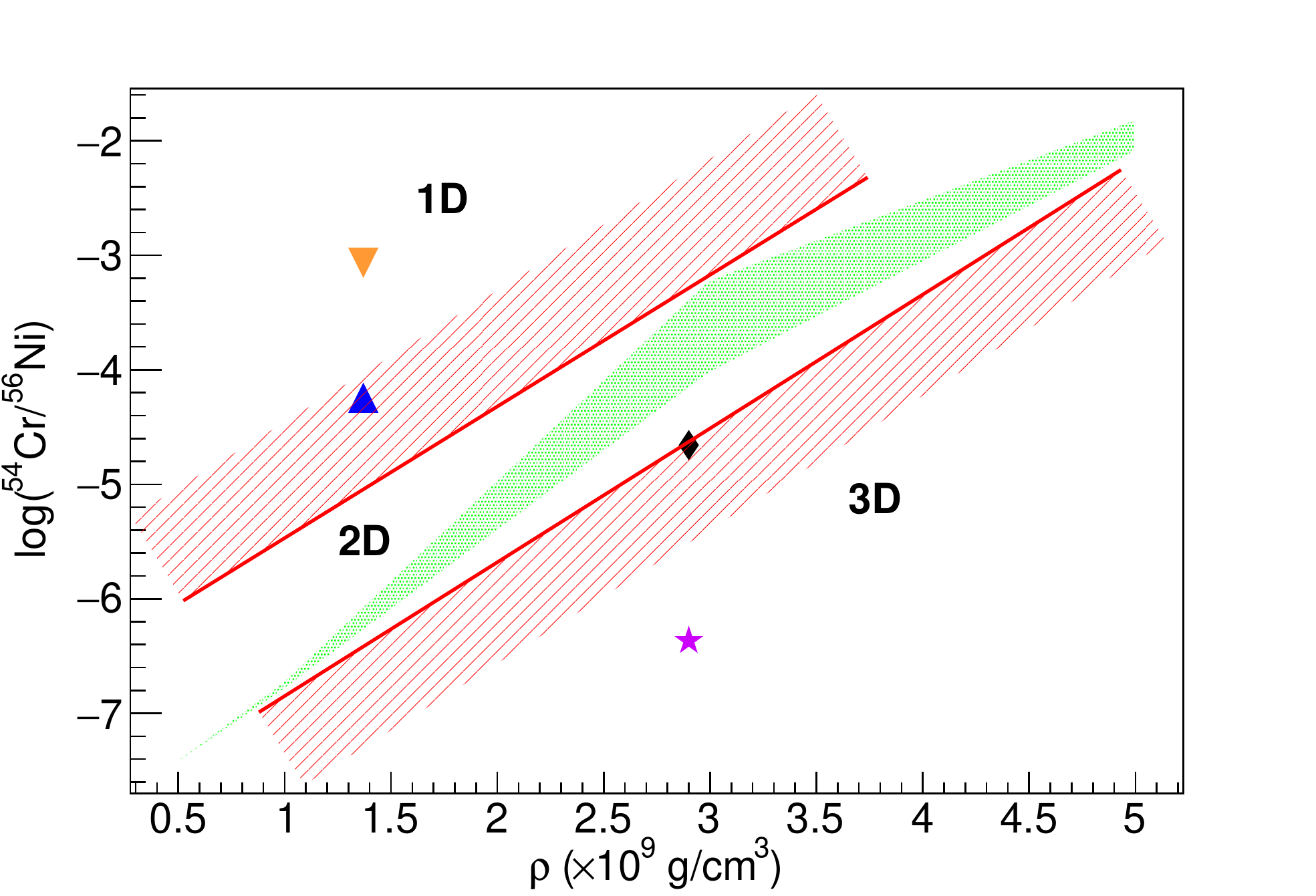}{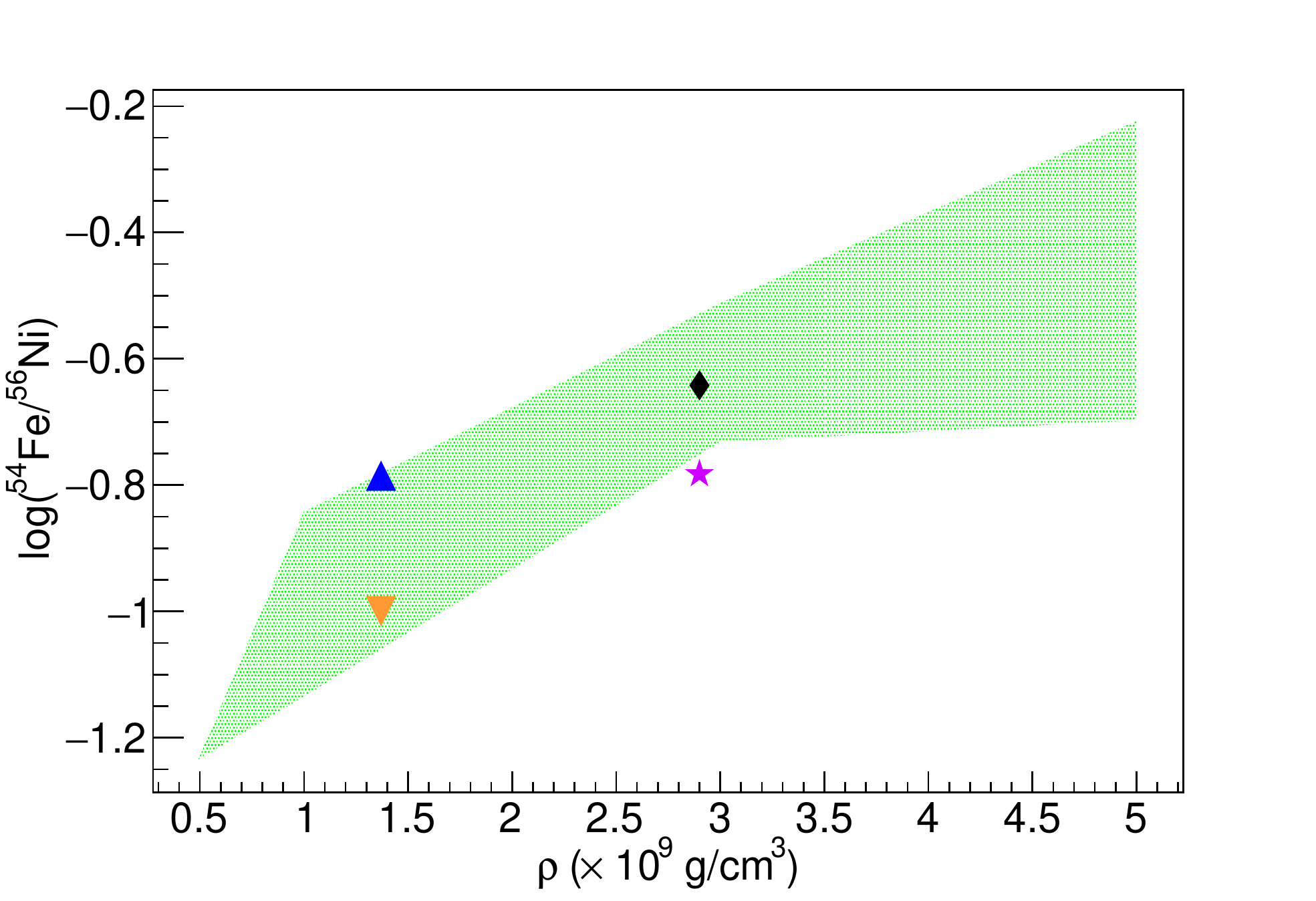}
\caption{\label{ej_mass}The ejected mass in the models normalized by the $^{56}$Ni mass. The green regions show the results of LN18. The red regions in the $^{54}$Cr case show expected density dependence for  {spherical}  and three dimensional models.}
\end{figure*}

This range of model uncertainty is illustrated in Figure \ref{ej_mass} where we compare the ejected mass ratios of  of $^{57}$Co/$^{56}$Ni, $^{55}$Fe/$^{56}$Ni, $^{54}$Cr/$^{56}$Ni, and $^{54}$Fe/$^{56}$Ni as a function of the WD central density. 
The green bands show the range of results of the delayed-detonation models in LN18.  So, for example the uncertainty in the $^{57}$Co/$^{56}$Ni ratio is taken from the highest and lowest ratio in the corresponding green band in {the} upper left panel of Fig.~\ref{ej_mass}.  This logarithmic range of uncertainty was then adopted for all of the models surveyed here.

Regarding Figure \ref{ej_mass}, there are a number of trends worthy of note.  
For one, The density dependence of $^{57}$Co production is not very strong in the {cyrindrically symmetric} models of LN18, while there appears to be a  positive correlation when going from the low density {spherical} W7 and WDD2 models to the 3D N100 and N150 models. 
For both $^{55}$Fe and $^{54}$Fe production, all models show such a correlation with density. 
Therefore the nucleosynthesis of $^{55}$Fe and $^{54}$Fe may be used as an diagnostic  of the density of the progenitor. 

The production of  $^{54}$Cr is particularly interesting.  Its yield spans five orders of magnitude, making its yield  the most sensitive  correlation with central density. 
The LN18 models show a strong positive correlation with density, while in other models the correlation may even be negative. 
We hypothesize that this  may reflect the {degree of symmetry} of the models as indicated by the schematic red bands in the corresponding lower left panel of Fig.~\ref{ej_mass}.   The {spherical W7 and WDD2 models} have a higher $^{54}$Cr production, 
while the three dimensional models have a lower production. 
We emphasize, however, that the tendency for one and three dimensional models (as shown by the red regions in this figure) may be indicative, but clearly need to be  substantiated with more simulations. 


We  also note  the differences between the deflagration and delayed-detonation models on this figure.
More $^{54}$Cr (with a $Y_e = 0.44$) is produced in WDD2 than in the W7 model.
However, the order is opposite for the production of $^{54}$Fe (with $Y_e = 0.48$).
Because each model is characterized by a different final $Y_e$,
a reversal in production is  perhaps expected, but this reversal is not as pronounced when comparing $^{56}$Ni ($Y_e=0.5$) and $^{57}$Co ($Y_e=0.47$). 
This indicates a lower $Y_e$ in the WDD2 model than in the W7 model.    

{However, this trend is reversed
for the 3D explosion calculations.  Given the results of  Figure \ref{ej_mass}, switching from the delayed
detonation model to the deflagration model results in a lower abundance of $^{54}$Fe, but higher abundance of $^{54}$Cr. 
In the case of 3D models, the deflagration model appears to result in a {lower} value for $Y_e$ as compared to the delayed detonation model.}

{Nevertheless}, there are some interesting trends among the models summarized in Table \ref{wd_calculations}.  Overall the sub-Chandra double-degenerate scenarios such as {the} violent merger model predict lower values for the $^{57}$Co/$^{56}$Ni and $^{55}$Fe/$^{57}$Co ratios, while the deflagration and Chandra single degenerate scenarios like the 2D LN18 models and the 3D N100 and N150def models predict {higher} ratios.    

A similar trend occurs in the Mn/Fe and Ni/Fe yields, i.e. that the sub-Chandra models predict a low Mn/Fe ratio while the Chandra deflagration models like N150def {predict} the highest ratio. These trends are compared with observations in the next section to possibly diagnose the nature of each observed SNIa explosion. 

\subsection{Comparison to observations}
In
Figures \ref{co_fe_ratio} and \ref{mn_fe_ratio}, we compare models to observations using the above elemental and isotopic yields as criteria.

\subsubsection{Comparison to light-curve data}
\begin{figure*}[htbp]
    \plottwo{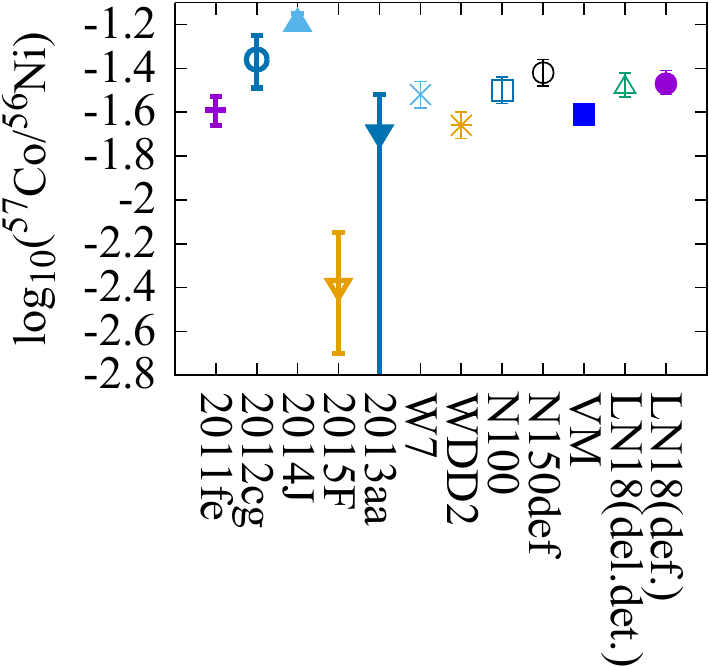}{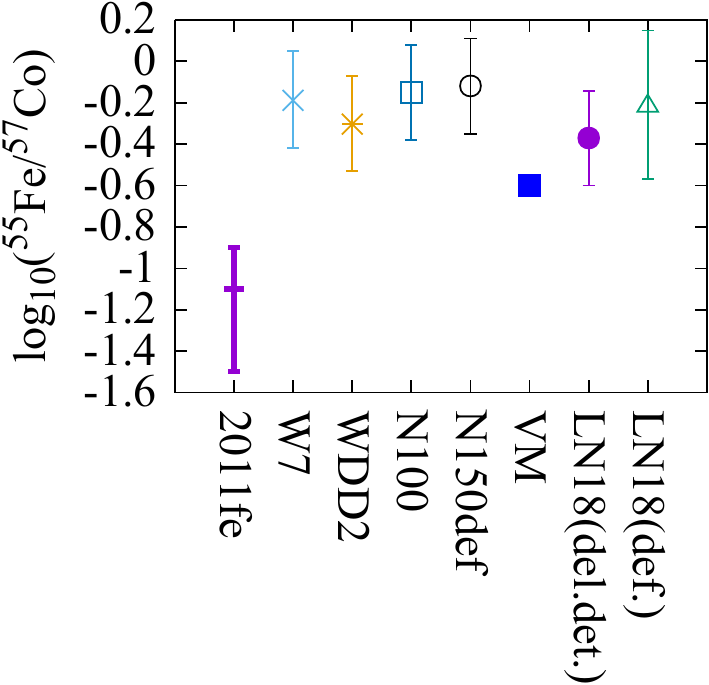}
    \caption{\label{co_fe_ratio}Ratios of the ejected mass. The thick points are the observational data and the thin points are the models. The variability of the delayed-detonation model results with model parameter variation as reported in \citet{leung18} (referred to as LN18 in the figure) is used as a guidance to {the} uncertainty of the theoretical model results.}
\end{figure*}

Figure \ref{co_fe_ratio} in particular shows observational results deduced from direct light-curve {SNIa} {photometry}. Observational data are shown as thick data points and error bars, while the thin data points with error bars are the yields of the theoretical models as labeled. 
As described  in the previous section we {indicate uncertainty in} the model results through 'error bars', which should reflect the variation of  yields due to the parameters of each model.

The left panel of Figure \ref{co_fe_ratio} shows the ratios of 
$^{57}$Co to $^{56}$Ni for the models from Table \ref{wd_calculations}
compared to observations of SN 2011fe, 2012cg, {2014J, 2015F, and 2013aa. The two events SN 2011fe and SN 2012cg compare well with the model predictions and suggest different explosion scenarios.} 

The predictions from the violent merger
calculation and the {spherical} delayed detonation (WDD2) model
are most consistent with observations of SN 2011fe, while the predictions
from the 3D deflagration model are most consistent with observations
of SN 2012cg. 
In both the 3D and 1D models, 
the production of $^{57}$Co relative to $^{56}$Ni is higher in the
deflagration model {than in} the delayed detonation model.
The {cylindrically symmetric} models of LN fall in between and appear consistent with both events.

{Recently, late light-curves of SN 2014J \citep{2014j}, 2015F \citep{2015f}, and 2013aa \citep{2013aa} have been measured and the $^{57}$Co/$^{56}$Ni ratios were estimated (Figure \ref{co_fe_ratio}). The result for SN 2013aa is just an upper limit but consistent with all of the models. The other two objects, however, cannot be explained by any of them. These results are subject to systematic uncertainties, so future observations are desireble to understand these events.}

The right side of Figure \ref{co_fe_ratio} shows the observed ratio of $^{55}$Fe to $^{57}$Co for SN 2011fe.  
All models produce somewhat higher ratios { [i.e.~$\log_{10}(^{55}$Fe$/^{57}$Co$)\sim -0.2$] than that deduced} from observation, although the violent merger model comes closest, within uncertainties.
The lower density associated with sub-$M_\mathrm{ch}$ models including the violent merger apparently results in this lower $^{55}$Fe/$^{57}$Co ratio.  

Based upon this figure we would identify SN 2011fe as most likely being  a sub-Chandra violent merger event, while SN 2012cg appears to be best fit with the higher-density single-degenerate Chandra deflagration models, particularly the highest-density N150def 3D model.

\begin{figure}[htbp]
    \plotone{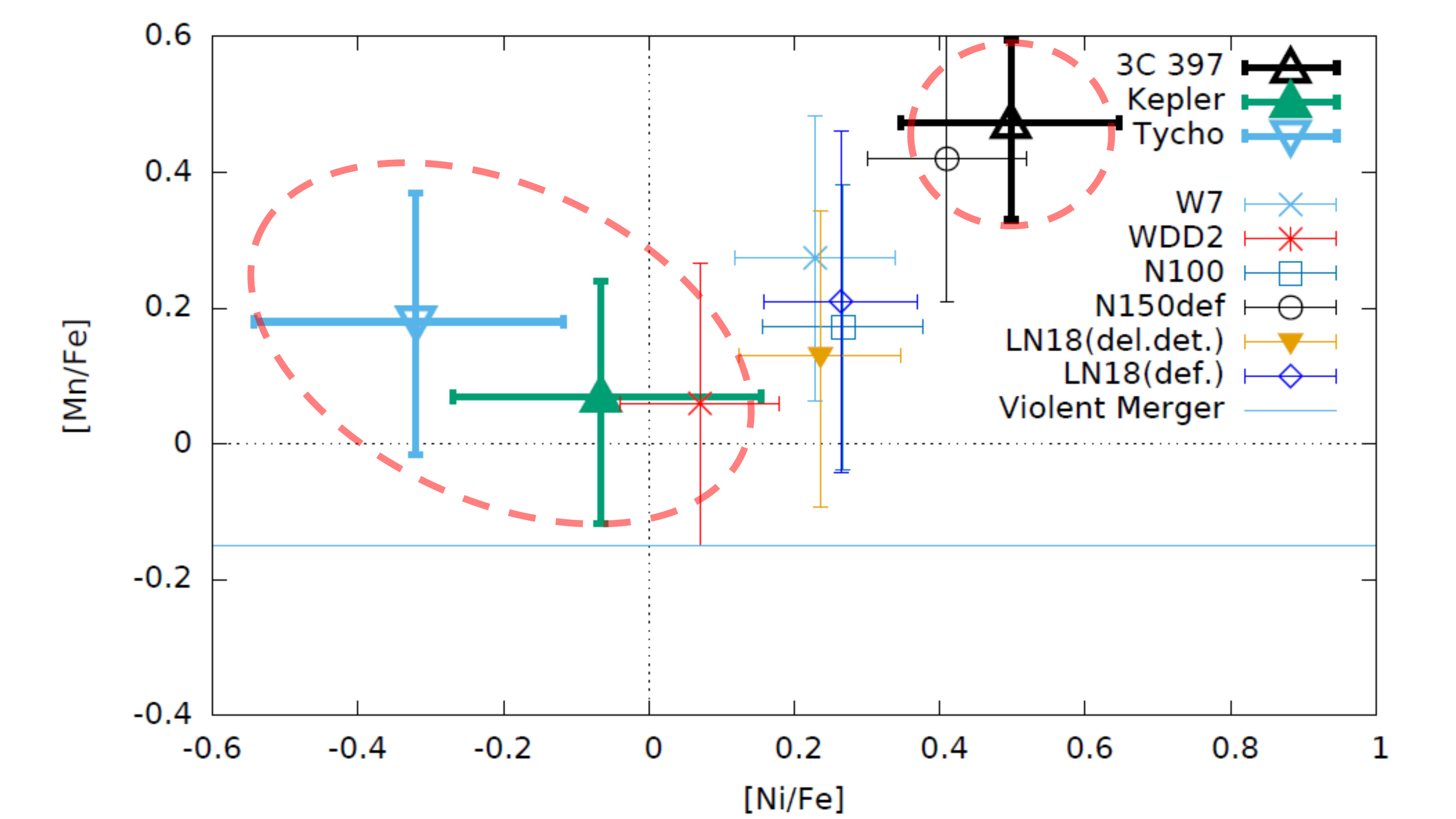}
    \caption{\label{mn_fe_ratio}Observational data of SNRs and theoretical results of SN Ia models. Dashed-oval regions are drawn to aid in identifying the groupings of data and models.  The thick points are the observational data and the thin points are the models. The variability of the delayed-detonation model results with model parameter as reported in \citet{leung18} is used as a guidance to uncertainty of the theoretical model results.}
\end{figure}

In Figure \ref{mn_fe_ratio} we show the correlation  from supernova remnant spectroscopy of the mass ratios of [Mn/Fe] and [Ni/Fe]. 
Three SN remnants,  3C 397, Kepler, and Tycho, are compared to the computational results from Table \ref{results}.  
Since no result for the [Ni/Fe] value is given from the violent merger model, the [Ni/Fe] value for this model  
is shown with a large  horizontal error bar  in the figure. {It is difficult to estimate uncertainties of [Mn/Fe] for the violent merger model, but any set of parameters in sub-Chandra models produces [Mn/Fe]$<0$ \citep{seitenzahl13b}.}

Although with only three remnants the number of observations is small,  [Mn/Fe] appears slightly above solar in all remnants.  
The dashed circles in this figure identify two {approximate}  groupings of  the models.  
The remnant 3C 397 shows much higher Mn and Ni relative to Fe and matches well the Chandra 3D N150def model.  
{The Tycho event, however, is not explained by any of the shown models. It is difficult to draw definitive conclusions on Kepler, but it is notable that Tycho and Kepler coincide with each other within the error bars. Therefore, they may originate from similar progenitors.} 

 Considering the uncertainties in the models, it is somewhat speculative to attempt to specify the explosion mechanisms
of these events.  Nevertheless, the difference between the 3C 397 and the overlapping results from the Kepler and Tycho supernova remnants may imply 
{that both explosion mechanisms occur},  {i.e. a Chandra single-degenerate explosion {higher-density} model for  3C 397 vs. a sub-Chandra  lower-density event for the Tycho and Kepler remnants.    }

With only one stable isotope ($A=55$) produced by the decay of $^{55}$Fe, the production of Mn may be more sensitive to the $Y_e$ in the explosion.  
The lower $Y_e$ of the deflagration model may result in a higher yield of Mn.  However, neither does the {spherical} deflagration model
 predict a convincingly-higher Mn yield, nor does the 3D delayed detonation model  predict the lower [Mn/Fe] and [Ni/Fe] values.


It is possible that continued spectral observations of supernova remnants can confirm this trend of  the explosion mechanisms of various progenitors. 
In particular, further observations of SNe Ia remnants may add data for  Mn and Ni with values in between those for the Kepler and 3C 397 remnants or that continue this trend of clumping around the sub-Chandra and Chandra models.


\begin{deluxetable*}{c|ccccccc}
\tablecaption{\label{wd_calculations} Summary of Explosion Models.$^*$}
\tablehead{Model& \colhead{W7[1]} & \colhead{WDD2[1]} &LN18 (del.det.)[6]&LN18 (def.)[6]& \colhead{N100[3]}&\colhead{N150def[4]}&\colhead{VM[2,5]}\\
Dimension&1&1&2&2&3&3&3\\
Mechanism$^{**}$ &Def.&Del. Det.&Del.~Det.&Def.&Del. Det.&Def.& Double Degen.\\
  & & & & & & & Sub-Chandra}
\startdata
$\log_{10}(^{57}\mathrm{Co}/^{56}\mathrm{Ni})$ & $-1.52$ & $-1.66$&-1.48&-1.47&$-1.50$&$-1.42$&$-1.61$ \\
$\log_{10}(^{55}\mathrm{Fe}/^{57}\mathrm{Co})$ & $-0.188$ & $-0.303$&-0.37&-0.21&$-0.150$&$-0.120$&$-0.601$ \\
$M_{^{56}\mathrm{Ni}}$ [$M_\odot$]& 0.651 & 0.668&0.530&0.300&0.604&0.378&0.616 \\
$M_{^{57}\mathrm{Co}}$ [$M_\odot$] & 0.0193 & 0.0144&0.0171&0.0106&0.0188&0.0143&0.0149\\
$M_{^{54}\mathrm{Cr}}$ [$M_\odot$]&$3.71\times 10^{-5}$&$5.77\times10^{-4}$&$2.40\times10^{-3}$&$2.42\times10^{-3}$&$2.61\times10^{-7}$&$8.28\times 10^{-6}$\\
$M_{^{54}\mathrm{Fe}}$ [$M_\odot$]&0.107&0.0670&0.107&0.101&0.0994&0.0862\\
Mn/Fe&0.0158&0.00964&0.0127&0.0160&0.0125&0.0219&0.00594\\
Ni/Fe&0.0916&0.0635&0.0960&0.102&0.0999&0.138\\
\enddata
\tablecomments{[1] \cite{mori16}; [2] \cite{roepke12}; [3] \cite{seitenzahl13}; [4] \cite{fink14}; [5] \cite{seitenzahl13b}; [6]\cite{leung18}}
\tablecomments{*Results from the
violent merger model do not report $^{54}$Cr or $^{54}$Fe yields. }
\tablecomments{** ``Def'' or ``Del.~Det.'' mean a deflagration model or a delayed-detonation Chandra model, respectively.}
 \end{deluxetable*}

\begin{deluxetable*}{c|cc|c|cc}
\tablecaption{\label{all_yields} 
Mass fractions for the nuclei produced in the W7 and WDD2 model. The {underlined nuclei  are produced in one model by more than twice as much as in the other model. The dashed-underlined  nuclei  are produced in the models at a level of} less than $10^{-5}M_\odot$.}
\tablehead{&\colhead{W7 [$M_\odot$]}&\colhead{WDD2 [$M_\odot$]}&  &\colhead{W7 [$M_\odot$]}&\colhead{WDD2 [$M_\odot$]}}
\startdata
\underline{$^{12}$C}& 4.794$\times 10^{-2}$&1.359$\times 10^{-3}$&$^{45}$Sc  & 1.639$\times 10^{-7}$&2.217$\times 10^{-7}$ \\ 
$^{13}$C  & 4.150$\times 10^{-8}$&3.292$\times 10^{-8}$&$^{46}$Ti&1.225$\times 10^{-5}$&1.321$\times 10^{-5}$ \\
\dashuline{ $^{14}$N } & 5.809$\times 10^{-6}$& 3.308$\times 10^{-8}$&$^{47}$Ti&4.021$\times 10^{-7}$&4.565$\times 10^{-7}$\\
\dashuline{ $^{15}$N } & 2.100$\times 10^{-8}$&5.473$\times 10^{-10}$&\underline{$^{48}$Ti} &2.385$\times 10^{-4}$&5.959$\times 10^{-4}$\\
$^{16}$O& 1.356$\times 10^{-1}$& 7.061$\times 10^{-2}$&\underline{$^{49}$Ti}&2.085$\times 10^{-5}$&4.195$\times 10^{-5}$\\
\dashuline{ $^{17}$O }  & 4.084$\times 10^{-6}$&6.553$\times 10^{-9 }$&\dashuline{ $^{50}$Ti }&2.184$\times 10^{-6}$&7.329$\times 10^{-5}$\\
\dashuline{ $^{18}$O } & 6.441$\times 10^{-8}$ &1.606$\times 10^{-10}$&$^{50}$V&4.721$\times 10^{-9}$&6.220$\times 10^{-9}$\\
\dashuline{ $^{19}$F }& 5.056$\times 10^{-10}$&4.061$\times 10^{-12 }$&$^{51}$V&8.376$\times 10^{-5}$&1.057$\times 10^{-4}$\\
$^{20}$Ne  & 1.309$\times 10^{-3}$& 1.072$\times 10^{-3}$&$^{50}$Cr&3.710$\times 10^{-4}$&3.868$\times 10^{-4}$\\
$^{21}$Ne  & 1.576$\times 10^{-6}$ &1.274$\times 10^{-6}$&$^{52}$Cr&7.209$\times 10^{-3}$&1.347$\times 10^{-2}$\\
\underline{$^{22}$Ne}  & 2.442$\times 10^{-3}$& 8.267$\times 10^{-4}$&$^{53}$Cr&1.184$\times 10^{-3}$&1.345$\times 10^{-3}$\\
\dashuline{ $^{23}$Na }  & 4.801$\times 10^{-5}$ &7.138$\times 10^{-6}$&\underline{$^{54}$Cr}&3.660$\times 10^{-5}$&5.737$\times 10^{-4}$\\
$^{24}$Mg  & 1.026$\times 10^{-2}$&7.081$\times 10^{-3 }$&$^{55}$Mn&1.259$\times 10^{-2}$&7.391$\times 10^{-3}$\\
\underline{$^{25}$Mg}  & 8.447$\times 10^{-5}$&2.943$\times 10^{-5}$ &$^{54}$Fe&1.077$\times 10^{-1}$&6.706$\times 10^{-2}$\\
$^{26}$Mg  & 4.478$\times 10^{-5}$ &3.215$\times 10^{-5}$&$^{56}$Fe&6.683$\times 10^{-1}$&6.834$\times 10^{-1}$\\
$^{27}$Al  & 8.509$\times 10^{-4}$&4.946$\times 10^{-4 }$&$^{57}$Fe&2.018$\times 10^{-2}$&1.503$\times 10^{-2}$\\
$^{28}$Si  & 1.732$\times 10^{-1}$&2.321$\times 10^{-1}$ &\underline{$^{58}$Fe}&1.741$\times 10^{-4}$&1.556$\times 10^{-3}$\\
$^{29}$Si & 7.921$\times 10^{-4}$ &4.786$\times 10^{-4}$&\underline{$^{59}$Co}&5.368$\times 10^{-4}$&1.905$\times 10^{-4}$\\
\underline{$^{30}$Si}  & 2.215$\times 10^{-3}$ &9.728$\times 10^{-4}$&$^{58}$Ni&6.936$\times 10^{-2}$&3.694$\times 10^{-2}$\\
$^{31}$P & 4.602$\times 10^{-4}$& 2.503$\times 10^{-4}$&$^{60}$Ni&4.523$\times 10^{-3}$&8.690$\times 10^{-3}$\\
$^{32}$S  & 7.890$\times 10^{-2}$ &1.317$\times 10^{-1}$&\underline{$^{61}$Ni}&6.557$\times 10^{-5}$&2.960$\times 10^{-4}$\\
$^{33}$S  & 3.312$\times 10^{-4}$ &2.272$\times 10^{-4}$&\underline{$^{62}$Ni}&6.165$\times 10^{-4}$&2.824$\times 10^{-3}$\\
$^{34}$S  & 1.712$\times 10^{-3}$ &2.291$\times 10^{-3}$& $^{64}$Ni &1.863$\times 10^{-6}$&2.322$\times 10^{-6}$\\
$^{36}$S  & 1.878$\times 10^{-7}$ &1.336$\times 10^{-7}$&\dashuline{ $^{63}$Cu }&2.499$\times 10^{-7}$&1.082$\times 10^{-6}$\\
$^{35}$Cl  & 1.045$\times 10^{-4}$ &8.783$\times 10^{-5}$&$^{65}$Cu&2.381$\times 10^{-7}$&3.616$\times 10^{-8}$\\
$^{37}$Cl  & 2.280$\times 10^{-5}$ &2.510$\times 10^{-5}$&\dashuline{ $^{64}$Zn }&1.749$\times 10^{-6}$&2.680$\times 10^{-5}$\\
$^{36}$Ar  & 1.324$\times 10^{-2}$ &2.519$\times 10^{-2}$&\dashuline{ $^{66}$Zn }&3.848$\times 10^{-6}$&4.184$\times 10^{-5}$\\
$^{38}$Ar  & 9.140$\times 10^{-4}$ &1.132$\times 10^{-3}$&\dashuline{ $^{67}$Zn }&2.445$\times 10^{-9}$&2.478$\times 10^{-8}$\\
$^{40}$Ar & 2.074$\times 10^{-9}$& 2.546$\times 10^{-9}$&\dashuline{ $^{68}$Zn }&1.313$\times 10^{-9}$&1.458$\times 10^{-8}$\\
$^{39}$K   & 6.752$\times 10^{-5}$ &6.611$\times 10^{-5}$\\
 $^{41}$K   & 4.068$\times 10^{-6}$ &5.026$\times 10^{-6}$\\
\underline{$^{40}$Ca}   & 1.133$\times 10^{-2}$& 2.477$\times 10^{-2}$\\
$^{42}$Ca   & 2.428$\times 10^{-5}$&2.901$\times 10^{-5 }$\\
$^{43}$Ca   & 7.151$\times 10^{-8}$&4.318$\times 10^{-8}$ \\
\dashuline{ $^{44}$Ca }   & 7.907$\times 10^{-6}$&2.675$\times 10^{-5}$ \\
 $^{46}$Ca  & 4.357$\times 10^{-10}$&4.885$\times 10^{-10}$ \\ 
\dashuline{ $^{48}$Ca }  & 2.778$\times 10^{-13}$& 1.965$\times 10^{-10}$\\ 
\enddata
\end{deluxetable*}

\section{Conclusions}
\label{discussion}
{In this work, we have {computed new} revised nucleosynthesis {yields} based upon new EC rates. We also made a new summary of the nucleosynthesis yields in {several recently analysed} light curves.} We have compared the results of various {spherical, cylindrical}, and 3D SNe Ia explosion
models in both single-degenerate Chandra models and double-degenerate sub-Chandra environments.  Simulations included deflagration and deflagration-plus-delayed-detonation explosions, 
and a 3D violent merger model.  Isotopic and
elemental yields for several key species were summarized for {these} models (Table \ref{wd_calculations}).
Elemental and isotopic ratios from these models were also compared to quantities inferred from SNIa  light curves and remnant observations.  For our comparisons, we attempt to include uncertainties in models from different parameter choices by adopting range of results from a recent broad parameter study with {cylindrically symmetric} models \citep{leung18}. 

 While no explosion model was able to reproduce 
the constraints of all of the observations, some models appear slightly better at 
reproducing different observational constraints simultaneously.
In particular, 
we find that the models and the observed nucleosynthesis tend divide  nearly equally into two groups roughly characterized by the central densities during thermonuclear burning.  One set of observations (including SN2011fe and the Kepler and Tycho remnants) is consistent with the low central densities of sub-Chandra double-degenerate merger models, while the other set (including SN2012cg and the 3C 397 remnant) favors the higher central densities of single-degenerate Chandra models.

{We note that observations of $^{54}$Cr and $^{54}$Fe
would be good diagnostics of the explosion model:  $^{54}$Cr is only produced in a significant abundance in the center of the WDD2 model.  
And the relative amounts of $^{54}$Cr and $^{54}$Fe apparently vary systematically with both
the explosion models (deflagration or delayed-detonation) and the deviation from spherical symmetry of
the explosion model.  A dependence on symmetry may come from the fact that spherically symmetric models tend to overestimate the degree of neutronization because of the suppressed buoyancy degrees of freedom.}

 Clearly there is a need for more observational elemental and isotopic ratios.  Only in this way can observations be tested for consistency with model predictions in all aspects of predicted light-curve shapes, elemental ratios around the iron group, and isotopic yields which are sensitive to the electron fraction and density of the burning regions. However,  a preliminary conclusion of the present study is a tendency for the observations to equally cluster around either the Chandra or sub-Chandra model results, suggesting that both models contribute substantially to observed SNIa events.
 
 {In this study, we used nucleosynthetic results from six Chandra models, while only the violent merger model is included as a sub-Chandra model. This is because detailed nucleosynthetic yields are calculated and published only for few sub-Chandra models. It is desirble that extended studies on nucleosynthesis in sub-Chandra models with large reaction networks be performed in the future.}
%

\acknowledgments
	KM's work was supported by NSF grant PHY-1430152;	MAF's by NSF grant PHY-1204486 and by
	an NAOJ Visiting Research Professorship; 
	TK's by JSPS KAKENHI Grant Numbers 
	JP15H03665 and JP17K05459; KN's and S.C.L.'s by the World Premier International Research Center Initiative
	(WPI), MEXT, Japan, and by JSPS KAKENHI Grant Numbers JP26400222, JP16H02168, and JP17K05382; and
	TS's was supported in part by  JSPS KAKENHI Grant number JP15K05090.
	The work of GJM is supported in part by the U.S. Department of Energy through Nuclear Theory Grant DE-FG02-95-ER40934.

\end{document}